\documentclass[10pt,letterpaper,twocolumn]{article} 

\usepackage{ol2}
\usepackage{hyperref}
\usepackage{amsmath}
\usepackage{amssymb}
\newcommand{\ud}{\mathrm{d}}

\begin{document}

\twocolumn[ 

\title{A collective modulation instability of multiple four-wave mixing}

\author{Andrea Armaroli and Stefano Trillo}
\address{
ENDIF, Department of Engineering, University of Ferrara, via Saragat 1, 44122 Ferrara, Italy\\
$^*$Corresponding author: andrea.armaroli@unife.it
}


\begin{abstract}
We investigate the modulation instability of multiple four-wave mixing arising from a dual-frequency pump in a single-mode fiber or waveguide.
By applying the Floquet theory on account of the periodic nature of four-wave mixing, we reveal a collective type of instability
occurring in the anomalous dispersion regime. Our interpretation of the linear stability analysis is validated 
by the numerical solution of the nonlinear Schr\"{o}dinger equation. 
\end{abstract}


 ] 

\maketitle
Two co-polarized pump frequencies $\omega_{0} \pm \Omega_d$ (detuning $2\Omega_d$) propagating in a Kerr medium produce multiple
four-wave mixing (FWM), i.e. generation of sideband pairs at $\omega_{0} \pm n \Omega_d$ with $n$ odd integer.
FWM is well known in fibers  \cite{Thompson91b,Hart94,Trillo94,Hart98},
and is being investigated also in other settings, e.g. semiconductor photonic crystal guides \cite{Eckhouse10}.
Despite long-lasting studies of FWM, a fundamental aspect such as the
stability of the process has been addressed only recently \cite{Trillo10}, 
though its impact on applications (e.g., regeneration \cite{Ciaramella01}, parametric amplification \cite{Mckinstrie02b}
and pulse-train generation \cite{Fatome06}) can be quite important.

In this Letter we investigate modulational instability (MI) of FWM. A well-known approach to MI of two pumps is based on
incoherently coupled nonlinear Schr\"{o}dinger (IC-NLS) equations, entailing
coupling only through cross-phase modulation (XPM) \cite{Agrawal87,Agrawal89}. 
While such model is valid for XPM-induced MI in appropriate polarization configurations \cite{Rothenberg90a,Millot02}, 
in the scalar case considered here (i.e., detuned pumps with parallel polarization), such model does not account for FWM.
Therefore the IC-NLS model has been argued to be inconsistent because the predicted MI
is resonant with FWM products $n=3$ ($\omega_{0} \pm 3\Omega_d$), neglected in the model \cite{Rothenberg90b}.

Here we show that the MI analysis can be carried out by accounting for FWM, if one adopts the technique originally developed in Refs. \cite{Trillo97b,Trillo97a}.
Using a more general model, namely a single NLS equation, which is known to account for multiple FWM \cite{Trillo94,Trillo10},
we describe FWM by use of four-mode truncation, assuming that the dynamics of higher-order FWM sidebands is locked to the first four lines
(the input dual-frequency pump and the first-order generated FWM sidebands). 
In this way, the recurrent (periodic) nature of FWM evolution \cite{Trillo94,Hart98} can be exploited to assess the stability of the process against
the growth of additional sideband pairs around the four waves by calculating the Floquet exponents of the linearized problem.
We interpret the gain curves calculated in this way and validate our interpretation by direct numerical integration of the NLS equation, 
finding a collective MI of FWM in the anomalous GVD regime.

Let us consider the dimensionless NLS equation for a slowly-varying envelope
$u(z,t)$ of a guided mode at carrier frequency $\omega_0$  
\begin{equation} \label{NLS}
i \frac{\partial u}{\partial z} - \frac{\beta}{2}\frac{\partial^2 u}{\partial t^2}+|u|^2u=0,
\end{equation}
where $\beta=k''/|k''|$ is the sign of group-velocity dispersion (GVD), $z=Z/Z_{nl}$ is the distance $Z$
in units of the nonlinear length $Z_{nl}=(\gamma P_{t})^{-1}$, and $t=(T-k' Z)/T_{s}$ stands
for the retarded time in units of  $T_{s}=\sqrt{Z_{nl} |k''|}$, $P_t$ is the total input power injected in the mode ($u$ is normalized so that it carries unit power),
and $\gamma=k_0 n_{2I}/A_{eff}$ is the standard nonlinear coefficient.

Following Ref.~\cite{Trillo94}, we describe the dynamics of FWM by means of four-mode truncation: 
\begin{equation}
\begin{split}
	u(z,t) = \frac{1}{\sqrt{2}} 
	[A_{p1}(z)\exp(i\Omega t) + A_{p2}(z)\exp(-i\Omega t)\\
	      + A_{s1}(z)\exp(i3\Omega t) + A_{s2}(z)\exp(-i3\Omega t) ],
\end{split}
\label{ansatz}
\end{equation}
where $A_{p1,p2}$ and $A_{s1,s2}$ are the cw complex amplitudes of $n=1$ (pumps) and $n=3$ (generated sidebands) FWM orders,
respectively, $2\Omega=2\Omega_d T_s$ being the normalized frequency detuning
(henceforth all frequencies denote actually normalized angular frequency offsets from $\omega_0$).
By substituting Eq.~\eqref{ansatz} in Eq.~\eqref{NLS} and collecting terms at frequencies $\pm\Omega,\,\, \pm3\Omega$, we obtain a set of
coupled-mode equations which govern the evolution of $A_{p1,s1}$:
\begin{equation}
\begin{split}
\begin{gathered}
	-i\frac{d{A}_{p1}}{dz}=
	\left( \frac{\left|A_{p1}\right|^2}{2}+ \left|A_{p2}\right|^2 + \left|A_{s1}\right|^2 +\left|A_{s2}\right|^2
	\right) A_{p1}  + \\	
	+A_{s1}A_{s2}A_{p2}^* + A_{s1}A_{p1}^*A_{p2} + \frac{A_{p2}^2A_{s2}^*}{2} + \frac{\beta \Omega^2}{2} A_{p1},  \\
	\label{FWM}
\end{gathered}\\
\begin{gathered}
	-i\frac{d{A}_{s1}}{dz} = 
	\left(\left|A_{p1}\right|^2 + \left|A_{p2}\right|^2 + \frac{\left|A_{s1}\right|^2}{2}+ \left|A_{s2}\right|^2 
	\right) A_{s1} + \\
	+\frac{A_{p1}^2A_{p2}^*}{2}+A_{p1}A_{p2}A_{s2}^* +  \frac{9\beta\Omega^2}{2} A_{s1},
\end{gathered}
\end{split}
\end{equation}
whereas $A_{p2,s2}$ obey Eqs.~\eqref{FWM} with subscripts $1$ and $2$ interchanged. 
The periodic solutions with period $z_p$ of Eqs.~\eqref{FWM} provide a good approximation to multiple FWM 
as long as $\Omega \gtrsim 1$ \cite{Trillo94}.
In order to investigate the MI of the periodic FWM process, we substitute in Eq.~\eqref{NLS}, a time-perturbed version of Eq.~\eqref{ansatz}, namely
\begin{equation}
\begin{aligned}
	A_k(z,t)=\left[ \sqrt{\eta_k(z)} + \varepsilon_k(z,t) \right] \exp \left(i \phi_k(z)\right), 
\end{aligned}
	\label{pertansatz}
\end{equation}
where $k=p1,p2,s1,s2$, $\eta_k$ and $\phi_k$ are, respectively, power and phase of the periodic solutions $A_k=\eta_k^{1/2} \exp(i \phi_k)$ of Eqs.~\eqref{FWM}, 
while we assume harmonic perturbations $\epsilon_k(z,t) = \varepsilon_{ka} \exp(-i \delta\omega t)+\varepsilon_{ks}\exp(i \delta \omega t)$,
corresponding to Stokes-antiStokes pairs at frequency offset $\delta \omega$ from each FWM component.
Retaining only linear terms in $\varepsilon_{ka,ks}$, we obtain the following system, which describes the linearized evolution
of the perturbation in a neighborhood of the dynamics ruled by Eqs.~\eqref{FWM}
\begin{equation}
\frac{\ud\mathbf{X}}{\ud z} = \mathbf{M}_p\left(\delta\omega\right)\mathbf{X},
\label{FloqSys}
\end{equation}
where $ \mathbf{X}(z) = ( \varepsilon_{p1s},\varepsilon_{p1a}^*,\varepsilon_{p2s},\varepsilon_{p2a}^*,
\varepsilon_{s1s},\varepsilon_{s1a}^*,\varepsilon_{s2s},\varepsilon_{s2a}^*)^T$ is the perturbation column array,
and  $\mathbf{M}_p(\delta \omega)$ is a complex 8x8 matrix, which depends on $\delta \omega$, and whose elements are
expressed in terms of periodic functions $\eta_k$ and $\phi_k$.
Since the matrix $\mathbf{M}_p(\delta \omega)$ is periodic with period $z_p$, Eq.~\eqref{FloqSys} lends itself to a Floquet analysis in order to assess its stability.
Floquet theory prescribes that the monodromy matrix of the system is constructed by integrating Eq.~\eqref{FloqSys} over the period 
with a set of linearly independent initial values (e.g.~the identity matrix). The eigenvalues $\lambda$ of such matrix are called Floquet multipliers: 
the perturbation is amplified exponentially with gain $g=2\log(|\lambda|)/z_p$, whenever  $|\lambda|>1$. 
We proceed to calculate the Floquet multipliers by jointly integrating over $z=[0,z_p]$ Eqs.~\eqref{FWM} and the linear system in Eq.~\eqref{FloqSys}.
Here we restrict ourselves to discuss symmetric FWM from initial conditions $A_{p1,p2}(0)=1$,
$A_{s1,s2}(0)=0$), reporting detailed results obtained for $\Omega=1.5$
as an illustrative example of the regime where FWM is non-negligible, while preserving the validity of the truncation ($\Omega \gtrsim 1$).
We also assume, for the time being, anomalous GVD.
\begin{figure}
\centerline{\includegraphics[width=7.5cm]{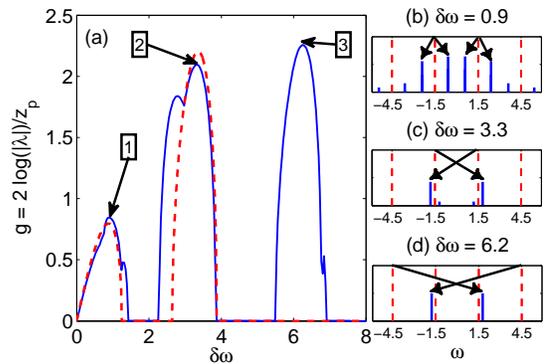}}
	\caption{(a) Gain curves from Floquet analysis (solid line) and IC-NLS (dashed line) vs.~MI frequency $\delta \omega$,
	for pump detuning $\Omega=1.5$ and anomalous GVD ($\beta=-1$). 
	(b-c-d) Structure of the unstable eigenvector at peak of bands 1 (b), 2 (c), and 3 (d). 
	The dashed red lines locate the FWM modes.
	The arrows connect the dominant MI sidebands  to the $A_k$ component they pertain to.}
	\label{f1}
\end{figure}
\begin{figure}[htb]
	\centerline{\includegraphics[width=9cm]{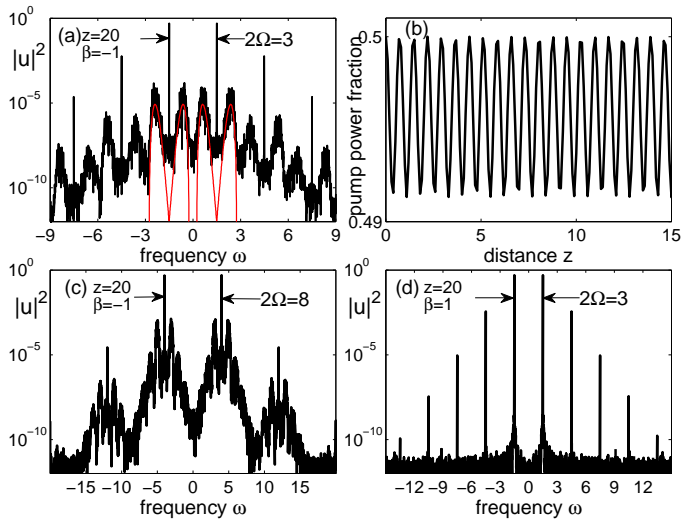}}
	\caption{(a,c,d) Output spectra from Eq.~\eqref{NLS}: anomalous GVD, $\Omega=1.5$ (a)
	and $\Omega=4$ (c). In (a) the gain band 1 around pumps are reported in thin solid red for comparison; 
	normal GVD, $\Omega=1.5$ (d); 
	(b) Periodic evolution ($z_p=0.71$) of pump power fraction relative to (a).}
	\label{f2}
\end{figure}

The gain curve reported in Fig.~\ref{f1}(a) shows the existence of three unstable branches labeled 1, 2, 3. 
Importantly, the first two branches are reminiscent of those due to XPM-MI (dashed curves) obtained from the IC-NLS model \cite{Agrawal87}.
Let us remind that, in particular, band 2 appears to be nearly resonant ($\delta \omega \sim 2\Omega$) with $n=3$ FWM products and, as such,
it was argued to be inconsistent with the IC-NLS model, where FWM is neglected \cite{Rothenberg90b}. 
Our result implies that such simple inconsistency argument cannot be invoked since the band survives when FWM is accounted for.
However,  by examining the composition of the unstable modes (i.e., the eigenvectors) in each of the three branches,
we reach the conclusion that, while branch 1 is effective for MI, branches 2 and 3 are fictitious in the sense that they do not give rise to different observable MI sidebands.
The reason is as follows. The unstable mode associated with branch 1, displayed in Fig.~\ref{f1}(b), implies the growth of nearly symmetric components 
around both pump waves, and weaker (nearly symmetric as well) components around FWM sidebands.
Viceversa the growing modes associated with band 2 and 3 turn out to have one pair of  strongly prominent components (out of the four possible pairs), 
namely $\varepsilon_{p1a}$, $\varepsilon_{p2s}$ (frequencies $\Omega - \delta \omega$ and $-\Omega + \delta \omega$) for band 2,
and  $\varepsilon_{s1a}$, $\varepsilon_{s2s}$ (frequencies $3\Omega - \delta \omega$ and $-3\Omega + \delta \omega$) for band 3.
In particular, consistently with the analysis based on IC-NLS, band 2 has vanishing quasi-resonant components $\varepsilon_{p2a},\varepsilon_{p1s}$,
while the strongest components $\varepsilon_{p1a}$, $\varepsilon_{p2s}$ are amplified thanks to the {\em single} decay process 
$\Omega + (-\Omega) \rightarrow (\Omega - \delta \omega) + (-\Omega + \delta \omega)$. In fact,
the peak frequency $\delta \omega=3.3$ is given by the positive root of the equation $\delta \omega (\delta \omega -2 \Omega) -1=0$, 
which is nothing but the nonlinear phase-matching condition of such process in normalized form. 
Given the peak frequencies of such branches 2 and 3 ($\delta \omega =3.3$ and $\delta \omega =6.2$, respectively)
the relative growing modes turn out to lie in the proximity of the pump frequencies [as sketched in Fig.~\ref{f1}(c-d)], 
thus overlapping with band 1. Photons at one of such frequencies, however, unavoidably lead, according to the full NLS dynamics,
to the generation of idler photons at mirror frequency, and both grow exponentially in pair (recall that they fall within band 1)
at the expense of a single pump, as entailed by Manley-Rowe relations of the dominant process [see Fig.~\ref{f1}(b)]. 
For this reason we expect MI due to branch 1 to always prevail against the other two branches.

This scenario has been validated by performing split-step simulations of Eq. \eqref{NLS},
using a two-pump input over a noisy background. The output spectrum at distance $z=20$, displayed  in Fig.~\ref{f2}(a) for $\Omega=1.5$,
clearly shows the growth of noise over MI bands which are symmetrically located around all FWM orders.
MI develops on top of the periodic exchange of energy between the pumps and FWM sidebands, as shown in Fig.~\ref{f2}(b).
Maximum amplification occurs at a modulation frequency $\delta \omega \simeq 0.9$ around all FWM orders,
which agrees well with the peak gain frequency of band 1 [superimposed  in Fig.~\ref{f2}(a)]. 
We find that the peak frequency $\delta \omega$ is nearly constant with pump detuning $\Omega$,
saturating to the value  $\delta \omega=1$ for $\Omega \gtrsim 3$. 
At such larger pump detunings also the growth of harmonics of $\delta \omega$ become visible [Fig.~\ref{f2}(c)].

\begin{figure}
	\centerline{\includegraphics[width=9cm]{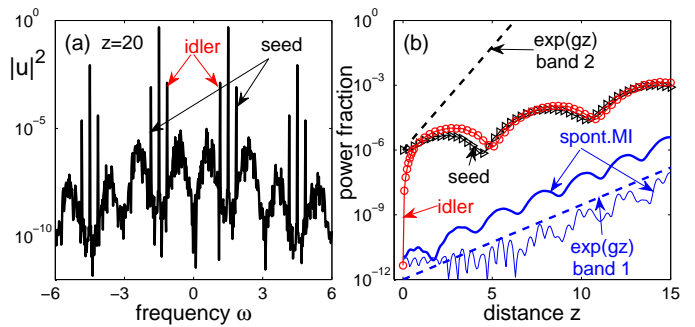}}
	\caption{(a) Output spectrum with additional input seed at $\delta \omega=3.3$ along the unstable eigenvector of band 2.
	(b) Power fraction evolution (in log scale) of seed (triangles), idlers (circles), 
	and peak of spontaneous MI growing from noise around the pump (thick solid line) and the $n=3$ FWM order (thin solid line). 
	The dashed lines give the predicted growth rates associated with peak gain of band 1 and 2 in Fig.~\ref{f1}, respectively.}
	\label{f3}
\end{figure}

To further confirm that neither band 2 nor band 3 leads to the exponential growth of asymmetric modes,
we have repeated the simulation of Fig.~\ref{f2}(a) with additional seeds along the unstable eigenvectors of Fig.~\ref{f1}(c-d). 
The case of band 2 is displayed in Fig.~\ref{f3}(a): 
besides spontaneous MI growing with the  rate predicted by band 1,  the output spectrum exhibits generated idlers with seed-idler pairs replicated around all FWM orders.
As it is clear from in Fig.~\ref{f3}(b) the idler is generated very rapidly from noise and then follows adiabatically the seed. 
Since the seed-idler pairs are then nearly symmetric and fall within MI-band 1, they are observed to grow (on average) with the gain characteristic of band 1 
(i.e. with a gain slightly lower than the peak gain of spontaneous MI) and not following the much higher growth rate
characteristic of band 2 (reported in Fig.~\ref{f3}(b) for comparison), which proves the correctness of our interpretation.

By extending our analysis also to the normal GVD regime, we have found that the low-frequency branch 1 responsible for MI disappears.
Our simulations of the NLS dynamics confirm that no MI occurs in this regime [see Fig.~\ref{f2}(d)].
Indeed the basic underlying mechanism of collective MI remains the decay of photon pairs from each pump
into sideband pairs [see Fig.~\ref{f1}(b)], which can be nonlinearly phase-matched only in the anomalous GVD regime. 
Such modulation is then transferred via the Kerr effect onto the generated FWM orders. This explains why the modulation frequency
$\delta \omega$ is fixed for all FWM orders and does not scale with the inverse square root of the relative power 
as would have been the case for a non-collective phenomenon ruled by conventional scalar MI.
Moreover, here MI sidebands grow  in an oscillatory way [Fig.~\ref{f3}(b)], 
owing to group-velocity mismatch between the pumps.

In summary, we have found that FWM exhibits a collective MI process that leads to the exponential growth of sidebands around all multiple FWM orders,
consistently with our Floquet analysis, once the latter is properly interpreted in terms of the associated eigenvectors.


\begin{thebibliography}{99}

\bibitem{Thompson91b} J. R. Thompson, and R. Roy,
``Nonlinear dynamics of multiple four-wave mixing processes
in a single-mode fiber,"
Phys. Rev. A {\bf 43}, 4987 (1991).

\bibitem{Hart94} D. L. Hart, A. F. Judy, T.A.B. Kennedy, R. Roy, and K. Stoev,
``Conservation law for multiple four-wave-mixing processes in a nonlinear optical medium,"
Phys. Rev. A {\bf 50}, 1807 (1994).

\bibitem{Trillo94} S. Trillo, S. Wabnitz, and T.A.B. Kennedy,
``Nonlinear dynamics of dual-frequency pumped multiwave mixing in optical fibers",
Phys. Rev. A {\bf 50}, 1732 (1994).

\bibitem{Hart98} D. L. Hart, A. F. Judy, R. Roy, and J. W. Beletic,
``Dynamical evolution of multiple four-wave mixing processes in an optical fiber,"
Phys. Rev. E {\bf 57}, 4757 (1998).

\bibitem{Eckhouse10}  V. Eckhouse, I. Cestier, G. Eisenstein, S. Combri\'{e}, P. Colman,  A. De Rossi,  M. Santagiustina, C. G. Someda,  and G. Vadal\`{a},
``Highly efficient four wave mixing in GaInP photonic crystal waveguides",
Opt. Lett. {\bf 35}, 1440 (2010).

\bibitem{Trillo10} S. Trillo and A. Valiani,
``Hydrodynamic instability of four-wave-mixing", 
Opt. Lett.  {\bf 35}, 3967 (2010).

\bibitem{Ciaramella01} 
E. Ciaramella, F. Curti, and S. Trillo, 
``All-optical signal reshaping by means of four-wave mixing in optical fibers",
IEEE Phot. Tech. Lett. {\bf 13}, 142 (2001).

\bibitem{Mckinstrie02b} C. J. McKinstrie, S. Radic, and A.R. Chraplyvy,
``Parametric amplifiers driven by two pump waves",
IEEE J. Sel. Top.  Quantum Electron. {\bf 8} 538-547 (2002).

\bibitem{Fatome06} J. Fatome, S. Pitois, and G. Millot, 
``20-GHz-to-1-Thz repetition rate pulse sources based on multiple four-wave-mixing in optical fibers",
\jqe {\bf 42}, 1038 (2006).

\bibitem{Agrawal87} G. P. Agrawal, ``Modulation instability induced by cross-phase modulation,"
Phys. Rev. Lett. {\bf 59}, 880 (1987).

\bibitem{Agrawal89} G. P. Agrawal, P. L. Baldeck, and R. R. Alfano, 
``Modulation instability induced by cross-phase modulation in optical fibers,"
Phys. Rev. A {\bf 39}, 3406 (1989).

\bibitem{Rothenberg90a} J. E. Rothenberg, 
``Modulational instability for normal dispersion," 
Phys. Rev. A {\bf 42}, R682 (1990).

\bibitem{Millot02} G. Millot, S. Pitois, and P. Tchofo-Dinda, 
``Modulational
instability processes in optical isotropic fibers under dual frequency
circular polarization pumping,Ó J. Opt. Soc. Am. B {\bf 19}, 454 (2002).

\bibitem{Rothenberg90b} J. E. Rothenberg, 
``Modulational instability of copropagating frequencies for normal dispersion," 
Phys. Rev. Lett. {\bf 64}, 813 (1990).

\bibitem{Trillo97b} S. Trillo and S. Wabnitz,
``Dynamic spontaneous fluorescence in parametric wave coupling,"
Phys. Rev. E {\bf 55}, R4897 (1997).

\bibitem{Trillo97a} S. Trillo and S. Wabnitz,
``Bloch wave theory of modulational polarization instabilities in birefringent optical fibers,"
Phys. Rev. E {\bf 56}, 1048 (1997).

\end{thebibliography}
\end{document}